\documentclass[11pt,a4paper]{article}

\usepackage[english]{babel}

\usepackage[T1]{fontenc}
\usepackage[math]{iwona} 

\usepackage{amsmath,amssymb}

\usepackage{graphicx}

\usepackage{placeins}

\usepackage{natbib} 
\usepackage{IEEEtrantools}

\usepackage{csquotes}

\newcommand{\E}[1]{\operatorname{\mathbb{E}}\!\left[#1\right]}

\usepackage{tikz} \usetikzlibrary{shapes}

\usepackage{hyperref} 

\usepackage{calc}

\usepackage[paper=a4paper,portrait, tmargin=30mm, bmargin=30mm, rmargin=40mm, lmargin=25mm]{geometry} 

\usepackage{tabls}

\newlength\micolumna
\newcommand{\figwidth}{0.6\columnwidth}

\title{Competitive advantage of URLLC vs. eMBB for supporting timeliness-relevant services\\

\thanks{This work was supported through Grant PID2021-123168NB-I00, funded by MCIN/AEI, Spain/10.13039/ 501100011033 and the European Union A way of making Europe/ERDF; Grant TED2021-131387B-I00, funded by MCIN/AEI, Spain/ 10.13039/501100011033 and the European Union NextGenerationEU/ RTRP; and by the Generalitat Valenciana, Spain, under Grant AICO/2021/138}}

\author{Luis Guijarro, José-Ramón Vidal, Vicent Pla\\
Universitat Politècnica de València}

\date{May 4, 2023}

\begin{document}

\thispagestyle{empty}

\begin{tabular*}{\textwidth}[t]{|p{\micolumna}|p{\micolumna}|}
	\hline
	\includegraphics[width=\micolumna,trim=2cm 17cm 2cm 2cm,clip]{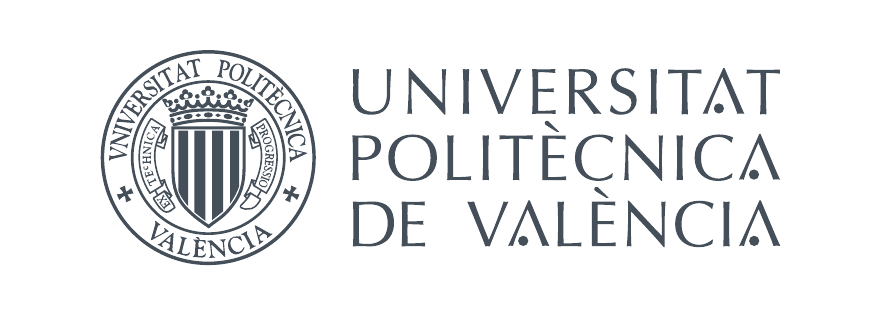}& 
	\raisebox{1.6cm}{\parbox{\micolumna}{Departamento de Comunicaciones}}\\
	\hline
	\multicolumn{2}{|c|}{}\\
	[0.20\textheight]
	\multicolumn{2}{|p{\textwidth-\tabcolsep*2}|}
	{\huge \begin{center}
Competitive advantage of URLLC vs. eMBB for supporting timeliness-relevant services
\end{center}}\\
	\multicolumn{2}{|c|}{\huge (Document NETECON129C)}\\
	[0.20\textheight]

	\hline
	Authors: Luis~Guijarro,  & Outreach: Public\\ 
	\cline{2-2}
	José Ramón Vidal, Vicent Pla  & Date: May 4, 2023 \\ 
	\cline{2-2}
	 & Version: b\\
	\hline
	\multicolumn{2}{|p{\textwidth-\tabcolsep*2}|}{Copyright notice: \enquote{This work has been submitted to the IEEE for possible publication. Copyright may be transferred without notice, after which this version may no longer be accessible.}}\\
	\hline
\end{tabular*}

\clearpage

\setcounter{page}{1}

\maketitle

\begin{abstract}
5G specifications promise a common and flexible-enough network infrastructure capable of satisfying diverse requirements of both current and future use cases. Two service types standardized in 5G are eMBB, without stringent delay guarantee, and URLLC, with stringent delay guarantee.
We focus on a use case where data timeliness is the relevant quality parameter.
We provide an economic rationale for the support of data-based services, that is, from the point of view of the profits attained by the service providers and operators (SP).
More specifically, we focus on data-based services the quality of which is related to the Age of Information, and we assess two alternatives for the support of this sort of services by means of a 5G network: one that is based on the eMBB service type, and one that is based on the URLLC service type. These assessment is conducted in a duopoly scenario.
We conclude that URLLC support provides a competitive advantage to an SP against a competitor SP that supports its service offering on eMBB. And that there is a slightly better situation for the users when the URLLC QoS constraint is stringent. 
\end{abstract}

\section{Introduction}

5G specifications promise a common and flexible-enough network infrastructure capable of satisfying diverse requirements of both current and future use cases. In order to guide the standandization process, three service types were initially defined~\citep{marsch2018}:
\begin{itemize}

\item Enhanced mobile broadband (eMBB), related to human-centric and enhanced access to multimedia content, services and data with improved performance and increasingly seamless user experience. This service type can be seen as an evolution of the services provided by 4G networks, and covers use cases with very different requirements, although the defining quality parameter is the data rate.

\item Ultra-reliable and low-latency communications (URLLC), related to use cases with stringent requirements for latency, reliability and availability. It is expected that URLLC services will provide a main part of the fundament of the Industry 4.0.

\item Massive machine-type communications (mMTC), capturing services that are characterized by a very large number of connected devices typically transmitting a relatively low volume of non-delay-sensitive data. However, the key challenge is that the devices are usually required to be low-cost, and have very long battery lifetime.
\end{itemize}

It should be stressed that many use cases envisioned in the 5G era cannot easily be mapped to one of the three main service types as listed above, as they combine the challenges and requirements related to multiple service types~\citep{popovski2021}. In the present work, we focus on a use case where data timeliness is the relevant quality parameter. In order to support this use case, we focus on two alternatives from the set of service types listed above: one without stringent delay guarantee, eMBB, and one with stringent delay guarantee, URLLC. We therefore factor out any quality requirement related to the device density, which is the relevant one in the design of mMTC service type.

The AoI is a performance indicator that is relevant for real-time status updating where the system has a source updating a monitor. The AoI describes the timeliness of a monitor's knowledge of the source. Although early works on its definition and analysis appeared more than a decade ago~\citep{kaul2012}, it has received a renewed interest within the context of MTC with stringent delay requirements such as the ones covered in this manuscript, but also encompassing tactile internet, edge cloud computing, and remote surgery systems~\citep{yates2021}.

The present work provides an economic rationale for the support of data-based services where timeliness is a relevant quality requirement by means of 5G networks.  More specifically, we focus on data-based services the quality of which is related to the Age of Information, and we assess two alternatives for the support of this sort of services by means of a 5G network: one that is based on the eMBB service type, and one that is based on the URLLC service type. These assessment is conducted in a duopoly scenario.

\begin{figure}[t]
\begin{center}
\includegraphics[width=\columnwidth]{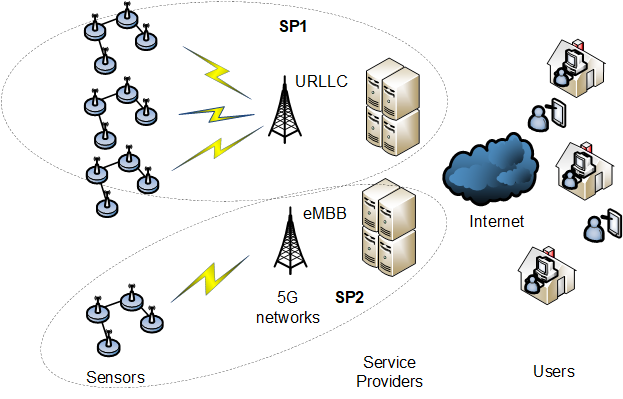}
\caption{Scenario}\label{fig:scenario}
\end{center}
\end{figure}

Fig.~\ref{fig:scenario} depicts the scenario under study. We assume that there are two Service Providers (SPs) that provide a service to users,  e.g., by means of an app. The users obtain a utility which depends on the timeliness of the data used by the service. Each SP uses a set of sensors that obtain the sensed data by means of a 5G network operated by the same SP: SP1's network provides a URLLC service, and SP2's network provides an eMBB service.

The structure of the manuscript is as follows. Section~\ref{sec:related} describes the related work. Section~\ref{sec:model} presents the system and economic models and states the analysis. Section~\ref{sec:results} presents and discusses the results and Section~\ref{sec:conclusions} provides the conclusions.

\section{Related work}\label{sec:related}

There are works that propose and analyze business models within the context of the Internet of Things. \citet{guijarro2011globecom,niyato2008} analyze communication services where the relevant quality parameter is the spectrum allocated by each operator to the pool of subscribers. \citet{guijarro2017jsac} analyses a service where the relevant quality parameter is the sensing rate of the data acquired by the service provider. \citet{guijarro2013wd,mandjes2003} analyze communication services where the relevant quality parameter is the delay incurred by the packets across the network. \citet{guijarro2021,caballero2019} analyze communication services over a 5G network implementing network slicing where  the quality parameter is the rate allocated by each tenant at each cell. However, none of them incorporates timeliness performance metrics into the user utility, as we do here.

On the other hand, the works that incorporate AoI in their analysis are not aimed at studying business models. These works are focused instead on the optimization of some AoI-based metric under different assumptions about the communication system responsible for the data delivery, e.g., the scheduling discipline of the queue model~\citep{kaul2012,kaul2012b}, the number of flows of status updates~\citep{yates2019,huang2015}, and the different priorities of the flows~\citep{kaul2018,najm2020}. In all these works, there is a bias towards the use of AoI as a performance metrics for a network service. This manuscript aims at incorporating the AoI into the utility that users derive from a service that runs over a network, specifically, from a data-based service where the timeliness of the data is a relevant quality parameter.

\section{Model}\label{sec:model}

We model a scenario (Fig.~\ref{fig:scenario}) where there are two SPs, each one receiving the data sensed by its sensors and communicated by means of a 5G network operated by the SP.  The data is used to provide a service to a set of users, which  derive a utility that depends on the timeliness of the data and are charged a subscription price.

\subsection{System model}

We model the 5G network as a queue of the $M/M/1$ type and a First-Come-First-Served (FCFS) discipline, which is dimensioned as described below.  That is, the network transfers the packets sent by the sensors to the SP, and is modeled as a one-server queue with Poisson packet arrivals with parameter $\lambda$ and independent exponentially distributed transmission duration with an expected value of $1/\mu$, where $\mu$ is the network capacity.

We measure the quality of the transfer of the sensor packets from the point of view of the SP service  by means of the AoI. The age of the information available at the monitor at time $t$ when the most recently received update is time-stamped $t_i$ is $\Delta(t)=t-t_i$. The average AoI can be obtained, under weak assumptions on the ergodicity of the service systems, as the limit of the time-averaged $\Delta(t)$.

It must be stressed that the goal of timely updating is neither the same as maximizing utilization of the communication system, nor as ensuring that generated status updates are received with minimum delay. Utilization may be maximized by making the source send updates as fast as possible. However, this may lead to the monitor receiving delayed statuses because the status messages become backlogged in the communication network. On the other hand, delay suffered by the stream of status updates can be minimized by reducing the rate of updates. However, reducing the update rate would lead to the monitor having excessively outdated status information because of a lack of updates. Minimizing an AoI metric is then a trade-off between these two objectives.

For an M/M/1-FCFS queue, the \textit{average AoI} is given by~\citet{kaul2012}:
\begin{equation}
\Delta = \frac{1}{\mu} \left( 1+\frac{1}{\rho}+\frac{\rho^2}{1-\rho} \right),
\end{equation}
where $\rho \equiv \frac{\lambda}{\mu}$.
There is an alternative AoI measure, the \textit{peak AoI,} which provides information about the worst case age, and is more tractable than the average AoI~\citep{costa2014}. For the M/M/1-FCFS queue, it is given by~\citet{huang2015}:
\begin{equation}
\Delta_p = \frac{1}{\mu} \left( 1+\frac{1}{\rho}+\frac{\rho}{1-\rho} \right)
\end{equation}

The dimensioning of the network capacity depends on the service class that is to be used to support the transmission of the sensing data:
\begin{itemize}
\item If an eMBB service is used, then the QoS requirement can be stated in terms of the average packet delay, say that it is below $\alpha$ times the minimum average delay: the delay that would see a packet sent over an unloaded network, which is $1/\mu$. In our queue model, this can be formulated as follows:
\begin{align}\label{eq:eMBB}
\frac{\E{t}}{1/\mu} & \leq \alpha\\
\lambda \leq \left( 1- \frac{1}{\alpha} \right) \mu
\end{align}
where $t$ is the system time in the queue model, which comprises both waiting and service times.

\item If a URLLC service is used, then a more restrictive performance criterion must be fulfilled: one that may involve a probabilistic measure over the packet delay across the network. In our M/M/1-FCFS queue model~\citep[p.~410]{stewart2009}, we will require that:
\begin{align}\label{eq:URLLCa}
\text{Prob}\{t > \epsilon \} &  \leq \delta,\\
e^{-\epsilon(\mu-\lambda)} & \leq \delta,\\
\lambda+\frac{1}{\epsilon} \ln \frac{1}{\delta} & \leq \mu, 
\end{align}

\end{itemize}

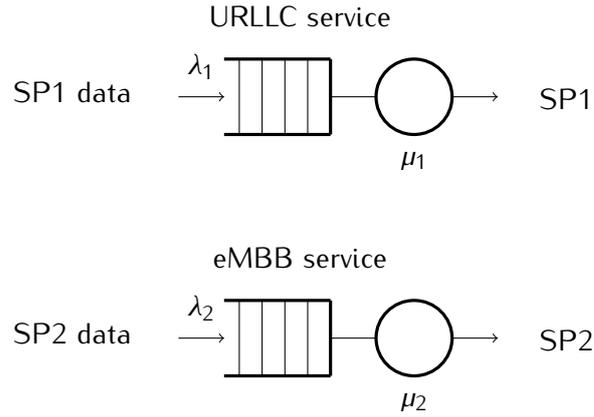
\begin{figure}[t]
\begin{center}
\begin{tikzpicture}

\draw[very thick](2,0)--(3.4,0);
\draw[very thick](2,1)--(3.4,1);
\draw(2.2,0)--(2.2,1);
\draw(2.5,0)--(2.5,1);
\draw(2.8,0)--(2.8,1);
\draw(3.1,0)--(3.1,1);
\draw[very thick](3.4,0)--(3.4,1);
\draw[very thick] (4.5,.5)circle(.5);
\draw(3.4,0.5)--(4,.5);
\draw[-To](1.4,0.5)--(2,.5);
\draw[-To](5,0.5)--(5.6,.5);
\node [above] at (4.5,-0.6) {$\mu_2$};
\node [below] at (1.7,1.2) {$\lambda_2$};
\node [above] at (3,1.3) {eMBB service};
\node [above] at (0,0.3) {SP2 data};
\node [above] at (6.5,0.2) {SP2};

\def\ya{3.2}

\draw[very thick](2,0+\ya)--(3.4,0+\ya);
\draw[very thick](2,1+\ya)--(3.4,1+\ya);
\draw(2.2,0+\ya)--(2.2,1+\ya);
\draw(2.5,0+\ya)--(2.5,1+\ya);
\draw(2.8,0+\ya)--(2.8,1+\ya);
\draw(3.1,0+\ya)--(3.1,1+\ya);
\draw[very thick](3.4,0+\ya)--(3.4,1+\ya);
\draw[very thick] (4.5,.5+\ya)circle(.5);
\draw(3.4,0.5+\ya)--(4,.5+\ya);
\draw[-To](1.4,0.5+\ya)--(2,.5+\ya);
\draw[-To](5,0.5+\ya)--(5.6,.5+\ya);
\node [above] at (4.5,-0.6+\ya) {$\mu_1$};
\node [below] at (1.7,1.2+\ya) {$\lambda_1$};
\node [above] at (3,1.3+\ya) {URLLC service};
\node [above] at (0,0.3+\ya) {SP1 data};
\node [above] at (6.5,0.2+\ya) {SP1};

\end{tikzpicture} 
\caption{System model}\label{fig:systemmodel}
\end{center}
\end{figure}

We focus on a scenario where one SP supports the delivery of the data by means of a URRLC service (SP1) and the other one, by means of an eMBB service (SP2). The system model is as depicted in Fig.~\ref{fig:systemmodel}.

\subsection{Economic model}

We assume that the two SPs compete against each other in order to provide a service to a population of $M$ users, which obtain a utility depending on the peak AoI of the data used to provide the service. Each user is charged a subscription price~$p$. 

\subsubsection{Users' subscription}

We model a set of $M$ users, which are homogeneous in their intrinsic value attached to the service provided by the SPs, a value that is modeled by a parameter $\nu$. 

Additionally, the users are heterogeneous in the intrinsic value that each one attaches to each SP, a value that can be related to those characteristics of the service offering not related to the network dimensioning, and therefore are modeled as an exogenous parameter $\gamma$, distributed uniformly in $[0,1]$. More specifically, we adopt a Hotelling model that accounts for the horizontal differentiation in the service offering, whereby a user with a value $\gamma$ suffers a disutility $\gamma$ when receiving service from SP1 and a disutility $1-\gamma$ when receiving service from SP2. Interestingly, $\gamma$ can be interpreted as a disutility for a user located at position $\gamma$ and SP1 and SP2 located at positions 0 and 1, respectively~\citep[p.~113]{belleflamme2015}.

The utility for a user indexed by $\gamma$ is proposed to be given by
\begin{equation}
u_1  = \nu + l/\Delta_{p1} - \gamma - p_1,
\end{equation}
if subscribes to SP1, and by
\begin{equation}
u_2  = \nu + l/\Delta_{p2} - (1-\gamma) - p_2,
\end{equation}
if subscribes to SP2.
Parameter $l$ translates quality into monetary units, which are the units for the other terms. The quality is given by the inverse of the peak AoI.

We assume, as argued below, that $p_1=p_2$. Furthermore, the value of the utility for the no-subscription option is zero and the intrinsic value $\nu$ is sufficiently high so that all $M$ users subscribe either to SP1 or SP2, that is, $u_1>0$ and $u_2>0$. In other words, the market is covered.
For given $p$'s, $\lambda$'s and $\mu$'s, the number of SP1 subscribers is given by:
\begin{align}\label{eq:users}
m_1 & = M\, \text{Prob}\{ u_1 \geq u_2 \}  \nonumber\\ 
& = M\, \text{Prob}\{ \gamma \leq \frac{1}{2}+\frac{l}{2}(\frac{1}{\Delta_{p1}}-\frac{1}{\Delta_{p2}}\}) \nonumber  \\
& = \begin{cases}
 0 & \text{if }   \Gamma < 0\\
 M \left( \frac{1}{2}+\frac{l}{2}(\frac{1}{\Delta_{p1}}-\frac{1}{\Delta_{p2}}) \right) & \text{if }  0  \leq \Gamma \leq 1\\
 M &  \text{if }  1  < \Gamma,
\end{cases}
\end{align}
where $\Gamma$ is defined as follows:
\begin{equation}
\Gamma \equiv \frac{1}{2}+\frac{l}{2}(\frac{1}{\Delta_{p1}}-\frac{1}{\Delta_{p2}}).
\end{equation}
The number of SP2 subscribers is given by $m_2= M - m_1$, under the assumption of complete market coverage.

Correspondingly,  the consumer surplus extracted by SP1 users can be computed as:
\begin{align}
CS_1 & = \int_{0}^{\min(\max(\Gamma,0),1)} (\nu + l/\Delta_{p1} - \gamma - p) d\gamma \\
& = \begin{cases}
 0 & \text{if }  \Gamma < 0\\
 \int_{0}^{\Gamma} (\nu + l/\Delta_{p1} - \gamma - p) d\gamma & \text{if } 0 \leq \Gamma \leq 1\\
 \int_{0}^{1} (\nu + l/\Delta_{p1} - \gamma - p) d\gamma &  \text{if } 1 < \Gamma.
\end{cases}
\end{align}
And the consumer surplus extracted by SP2 users can be computed as:
\begin{align}
CS_2 & = \int_{\min(\max(\Gamma,0),1)}^{1} (\nu + l/\Delta_{p2} - \gamma - p) d\gamma \\
& = \begin{cases}
 \int_{0}^{1} (\nu + l/\Delta_{p2} - (1-\gamma) - p) d\gamma & \text{if }  \Gamma < 0\\
 \int_{\Gamma}^{1} (\nu + l/\Delta_{p2} - (1-\gamma) - p) d\gamma & \text{if } 0 \leq \Gamma \leq 1\\
 0 &  \text{if } 1 < \Gamma.
\end{cases}
\end{align}
The consumer surplus aggregates the utility of all users, and the expressions above assume complete market coverage.

\subsubsection{Service provider's decisions}

We focus on the investment and procurement decisions to be taken by each SP in order to maximize its profits, that is, the network capacity (investment) and the sensor traffic (procurement). And we defer the pricing decision to a further study, so that the service price is assumed to be set, and therefore a parameter. For simplicity, furthermore, prices are assumed identical.

SPi's profit $\Pi_i$ is equal to the service revenue minus the costs. The revenue is given by $ m_i p$, and the relevant costs for our analysis are the investment costs, which are modeled as an increasing and convex function of the network capacity $\mu_i$, specifically, a quadratic cost function $c \mu_i^2$. The variable costs due to the packet transmission are assumed negligible, and therefore zero. Thus:
\begin{equation}
\Pi_i = m_i p -c \mu_i^2,\qquad i=1,2.
\end{equation}

The investment variable is $\mu_i$ and the procurement variable is $\lambda_i$, which influence the service quality. That is, the profit maximizing problem that is faced by each SP can be formulated as follows:
\begin{equation}\label{eq:PMP}
\underset{\mu_i,\lambda_i}{\max}\qquad  \Pi_i ,\qquad i=1,2.
\end{equation} 

As can be observed in~\eqref{eq:users}, there is a dependence of profit $\Pi_1$ not only on SP1's decision variables $\mu_1$ and $\lambda_1$, but also on SP2's decison variables $\mu_2$ and $\lambda_2$, and viceversa. This interaction between SP1 and SP2 is of an strategic nature, so that the analysis of this two linked maximization problems is best suited to Game Theory.


Incorporating the dimensioning constraints put by the eMBB/URLLC support~\eqref{eq:eMBB},\eqref{eq:URLLCa}, the two profit maximization problems~\eqref{eq:PMP} are the following ones: 

\begin{IEEEeqnarray} {rCCl}\label{eq:URLLCPMP}
 \underset{\mu_1,\lambda_1}{\max} & & \Pi_1 \\
 \text{subject to } & &  \lambda_1+\frac{1}{\epsilon} \ln \frac{1}{\delta} & \leq \mu_1. \nonumber
\end{IEEEeqnarray} 

\begin{IEEEeqnarray} {rCCl}\label{eq:eMBBPMP}
 \underset{\mu_2,\lambda_2}{\max } & & \Pi_2 \\
 \text{subject to } & & \lambda_2 \leq \left( 1- \frac{1}{\alpha} \right) \mu_2,	\nonumber
\end{IEEEeqnarray} 

The set of values $\mu_1^*$, $\lambda_1^*$, $\mu_2^*$ and $\lambda_2^*$ that simultaneously solve the above two problems is known as the Nash equilibrium of the game, and it is the most common equilibrium concept used in economics for modeling a scenario where few firms compete against each other in a strategic setting. 
At a Nash equilibrium $(\mu_1^*,\lambda_1^*,\mu_2^*,\lambda_2^*)$ no player (say SP1) has an incentive to deviate to a different strategy (say $(\mu_1^{\prime},\lambda_1^{\prime})$) provided that the other players (i.e., SP2) stick to the equilibrium strategy (i.e.,$(\mu_2^*,\lambda_2^*)$). In other words, SP1's profit does not improve through deviation (i.e., $\Pi_1(\mu_1^*,\lambda_1^*,\mu_2^*,\lambda_2^*) \geq \Pi_1(\mu_1^{\prime},\lambda_1^{\prime},\mu_2^*,\lambda_2^*) \; \forall \mu_1^{\prime},\lambda_1^{\prime}$). And the same for SP2.

\section{Results}\label{sec:results}

We discuss the results in terms of network dimensioning (Fig.~\ref{fig:dimension-comp}), AoI (Fig.~\ref{fig:AoI-comp}), number of users (Fig.~\ref{fig:users-comp}), consumer surplus (Fig.~\ref{fig:CS-comp}) and profit (Fig.~\ref{fig:profits-comp}) for the competing SPs, each one with a different support alternatives in the 5G network: URLLC (SP1) and eMBB (SP2). And in terms of consumer surplus (Fig.~\ref{fig:CS-comp}), total profit (Fig.~\ref{fig:profits-comp}) and social welfare (Fig.~\ref{fig:SW-tot}) for the aggregate.

We conduct below comparative statics, that is, we characterize the different Nash equilibria that result as one parameter is varied across a range of values. Specifically, we conduct two comparative statics. First, we analyze the effect of parameter $\epsilon$, which characterizes how stringent the QoS requirement for URLLC is; and second, we analyze the effect of parameter $c$, which characterizes the investment cost of the network.

The Nash equilibria were obtained by solving numerically the maximization problems~\eqref{eq:URLLCPMP} and~\eqref{eq:eMBBPMP}.

The parameters used are $M=10$, $c=0.1$, $l=0.5$, $p=1$, $\nu=2$, $\alpha=3$, $\epsilon=0.8$ and $\delta=0.1$, if not stated otherwise.  

\subsection{Effect of parameter \texorpdfstring{$\epsilon$}{epsilon}}

Here, the parameter $\epsilon$ varies between 0.3 and 2.0.

The results show that as $\epsilon$ increases,  both eMBB and URLLC support become equivalent. This is so because the URLLC constraint~\eqref{eq:URLLCa} is no longer more stringent than the eMBB constraint~\eqref{eq:eMBB}, so that there is no difference between~\eqref{eq:eMBBPMP} and~\eqref{eq:URLLCPMP}.

\begin{figure}[t]
\begin{center}
\includegraphics[width=\figwidth]{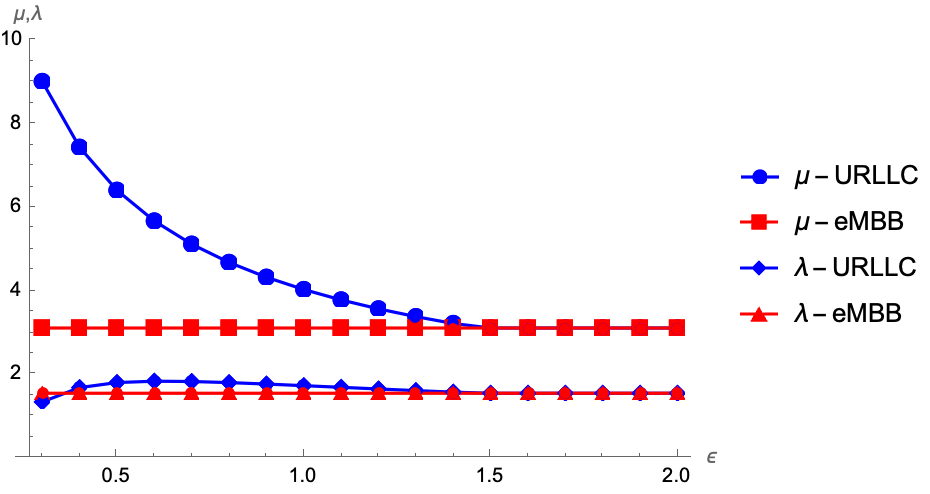}
\caption{$\mu$ and $\lambda$ as functions of $\epsilon$ for URLLC SP and eMBB SP}\label{fig:dimension-comp}
\end{center}
\end{figure} 

Fig.~\ref{fig:dimension-comp} shows that the SP with URLLC support chooses a lower network utilization (i.e., the ratio $\lambda/\mu$) than the eMBB SP, which is consistent with the fact that the quality requirements are more stringent in URLLC than in eMBB.

\begin{figure}[t]
\begin{center}
\includegraphics[width=\figwidth]{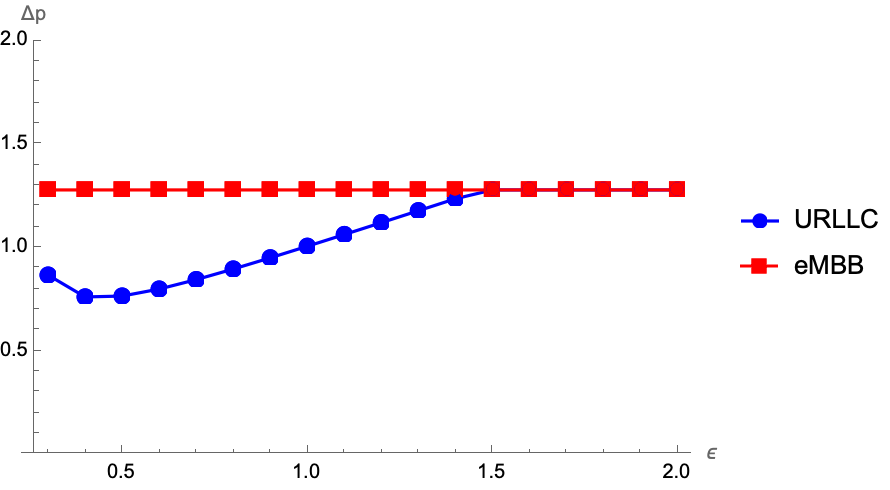}
\caption{AoI as a function of $\epsilon$ for URLLC SP and eMBB SP}\label{fig:AoI-comp}
\end{center}
\end{figure} 

Fig.~\ref{fig:AoI-comp} shows that URLLC support achieves a better (lower) AoI than eMBB support, so that the lower network utilization that URLLC requires translates into a better quality for the service than eMBB.

\begin{figure}[t]
\begin{center}
\includegraphics[width=\figwidth]{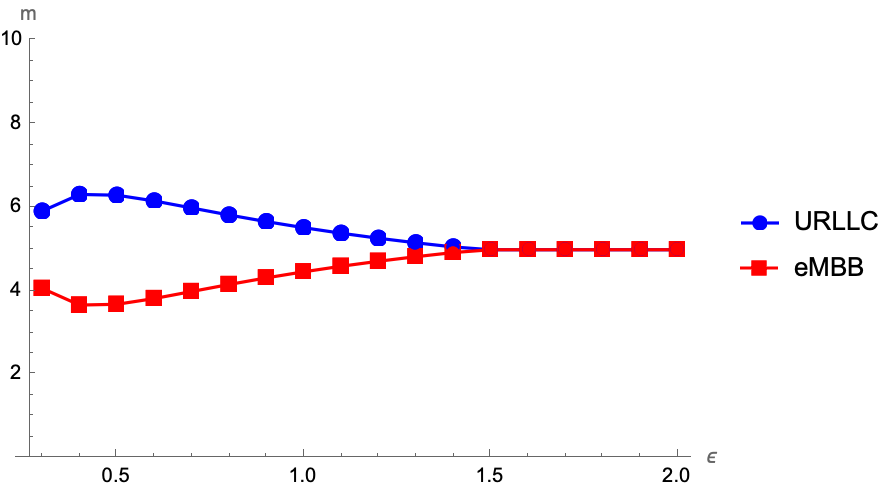}
\caption{Number of subscribers as a function of $\epsilon$ for URLLC SP and eMBB SP}\label{fig:users-comp}
\end{center}
\end{figure} 

Fig.~\ref{fig:users-comp} shows that URLLC support achieves a higher subscription base than eMBB support, which is consistent with the AoI results.

\begin{figure}[t]
\begin{center}
\includegraphics[width=\figwidth]{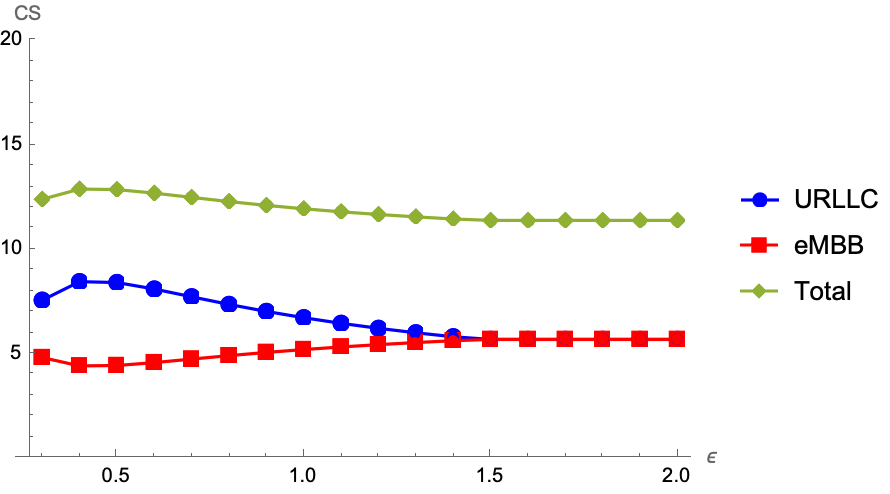}
\caption{Consumer surplus as a function of $\epsilon$ for URLLC SP and eMBB SP and total consumer surplus}\label{fig:CS-comp}
\end{center}
\end{figure} 

\begin{figure}[t]
\begin{center}
\includegraphics[width=\figwidth]{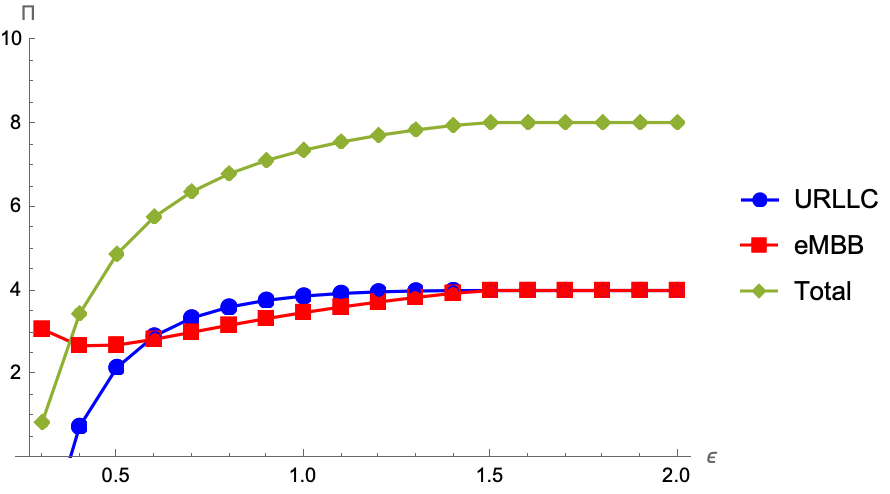}
\caption{SPs' profits and total profit as a function of $\epsilon$}\label{fig:profits-comp}
\end{center}
\end{figure}

Fig.~\ref{fig:CS-comp} shows that the aggregated surplus of the URLLC-supported users is greater than the surplus of the eMBB-supported users, which is sensible since it is a measure that aggregates the quality of the service, the price and the number of subscribers. The same figure shows that the total consumer surplus does not exhibit high variation with $\epsilon$, but it peaks when the number of URLLC-supported users is maximum and the AoI is minimum. 

Fig.~\ref{fig:profits-comp} shows that URLLC support provides higher profits for not so stringent $\epsilon$. For the parameters chosen, it may however cause losses if the URLLC quality constraint is very stringent, i.e. very low $\epsilon$. It also shows that the aggregated profit increases significantly with $\epsilon$.

\begin{figure}[t]
\begin{center}
\includegraphics[width=\figwidth]{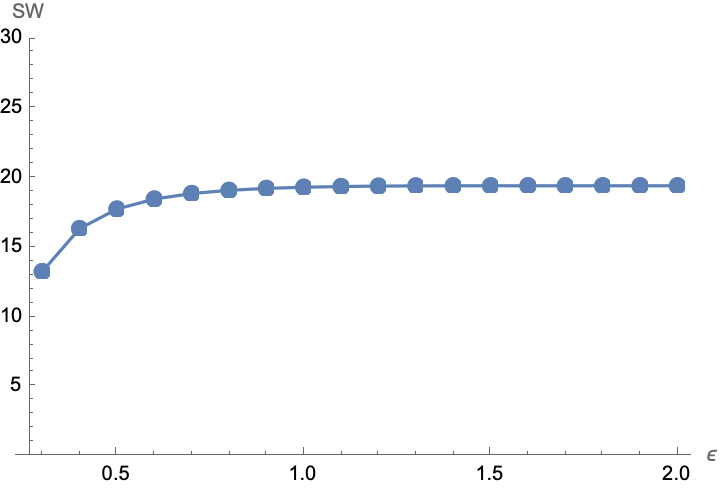}
\caption{Social welfare as a function of $\epsilon$}\label{fig:SW-tot}
\end{center}
\end{figure} 

Finally, when aggregating the SPs' profits and the consumer surplus into the social welfare quantity, Fig.~\ref{fig:SW-tot} shows that the variation of the aggregated profit dominates.

\FloatBarrier

\subsection{Effect of parameter \texorpdfstring{$c$}{c}}

Here, the parameter $c$ varies between 0.08 and 0.4.

\begin{figure}[t]
\begin{center}
\includegraphics[width=\figwidth]{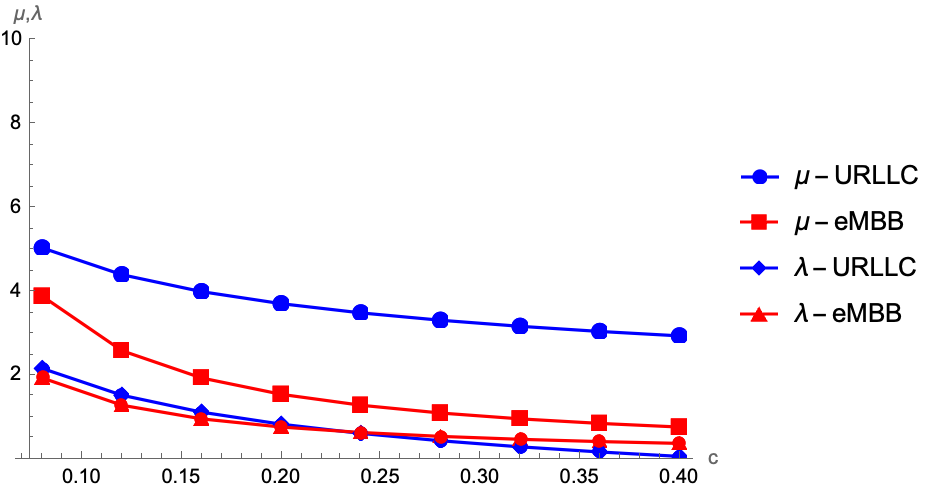}
\caption{$\mu$ and $\lambda$ as functions of $c$ for URLLC SP and eMBB SP}\label{fig:dimension-comp2}
\end{center}
\end{figure} 

Fig.~\ref{fig:dimension-comp2} shows that the SP with URLLC support chooses again a lower network utilization than the eMBB SP. Furthermore, as $c$ increases, the network deployment becomes more expensive and both SPs reduce their capacity.

\begin{figure}[t]
\begin{center}
\includegraphics[width=\figwidth]{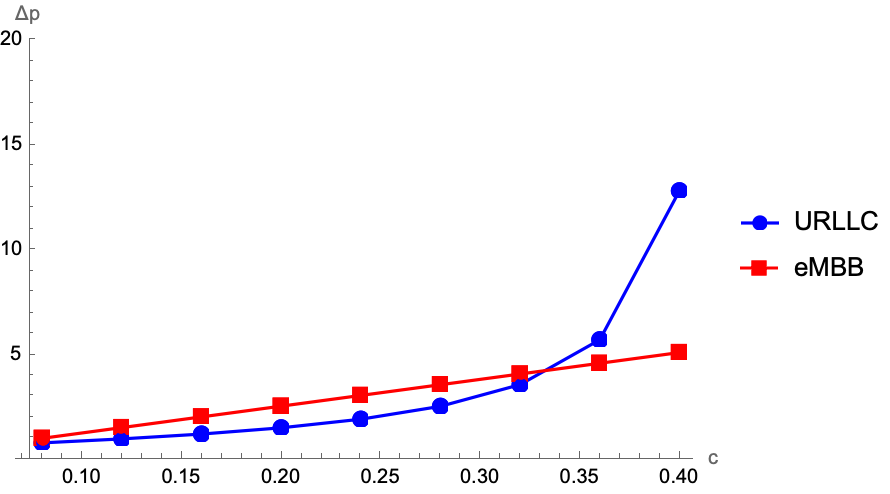}
\caption{AoI as a function of $c$ for URLLC SP and eMBB SP}\label{fig:AoI-comp2}
\end{center}
\end{figure} 

Fig.~\ref{fig:AoI-comp2} shows that URLLC support achieves a better (lower) AoI than eMBB support, as long as the network deployment is not very expensive.

\begin{figure}[t]
\begin{center}
\includegraphics[width=\figwidth]{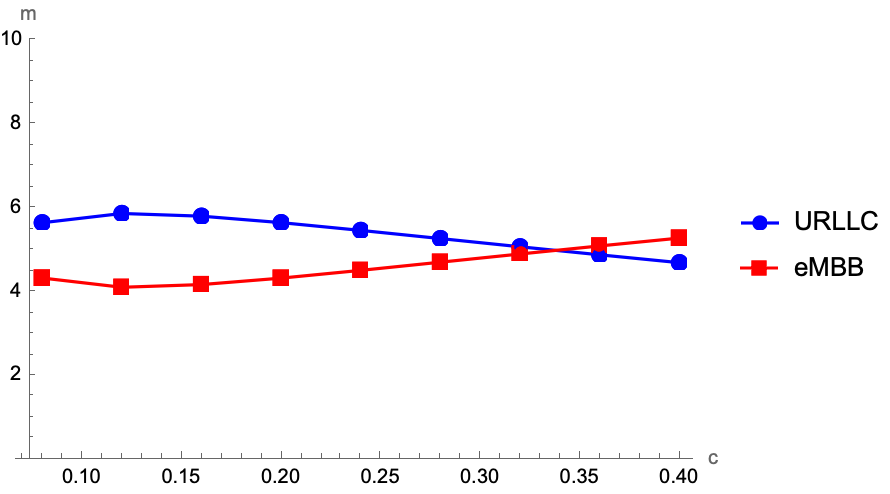}
\caption{Number of subscribers as a function of $c$ for URLLC SP and eMBB SP}\label{fig:users-comp2}
\end{center}
\end{figure} 

Fig.~\ref{fig:users-comp2} shows that URLLC support achieves a higher subscription base than eMBB support, which is consistent with the AoI results, again for not very expensive network deployment.

\begin{figure}[t]
\begin{center}
\includegraphics[width=\figwidth]{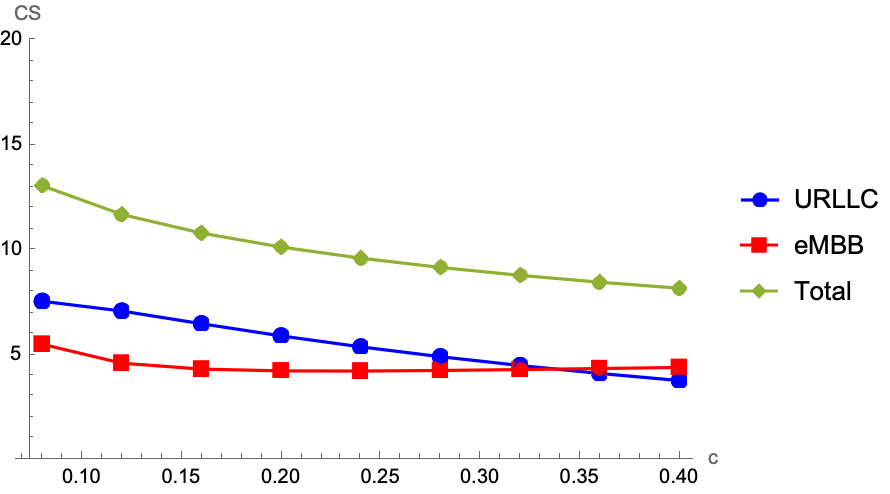}
\caption{Consumer surplus as a function of $c$ for URLLC SP and eMBB SP and total consumer surplus}\label{fig:CS-comp2}
\end{center}
\end{figure} 

\begin{figure}[t]
\begin{center}
\includegraphics[width=\figwidth]{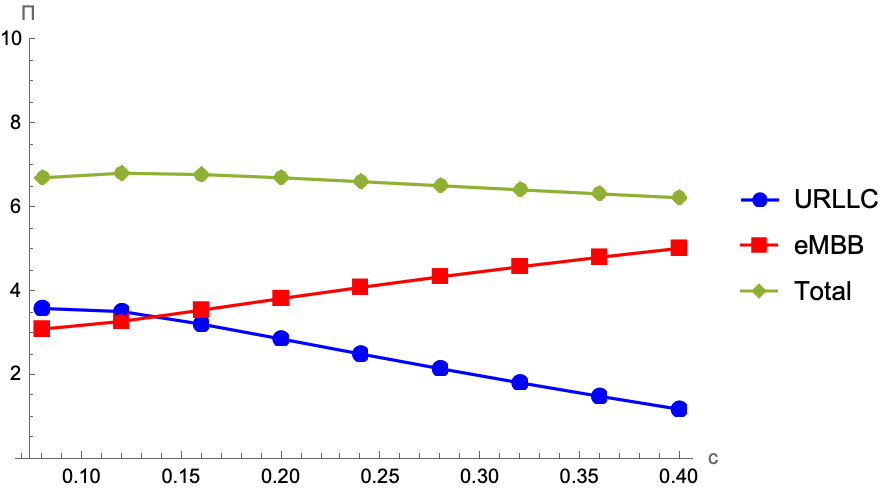}
\caption{SPs' profits and total profit as a function of $c$}\label{fig:profits-comp2}
\end{center}
\end{figure}

Fig.~\ref{fig:CS-comp2} shows that the aggregated surplus of the URLLC-supported users is greater than the surplus of the eMBB-supported users. The same figure shows that the total consumer surplus decreases as $c$ increases. 

Fig.~\ref{fig:profits-comp2} shows that URLLC support provides lower profits than eMBB support except for very low $c$. The sum of profits only decreases slowly as $c$ increases.

\begin{figure}[t]
\begin{center}
\includegraphics[width=\figwidth]{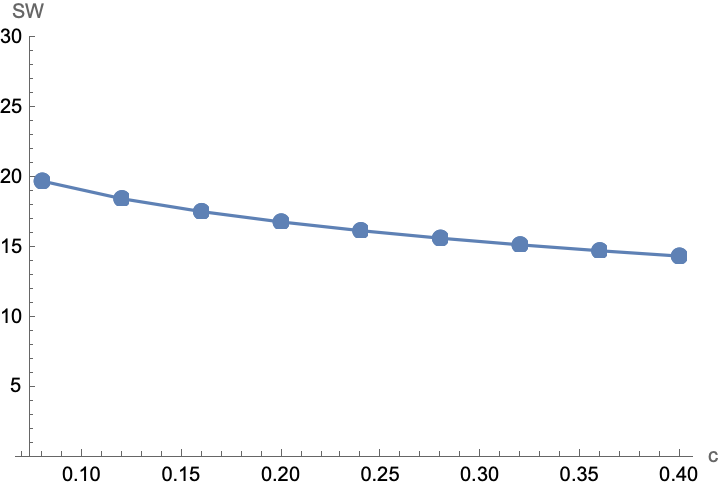}
\caption{Social welfare as a function of $c$}\label{fig:SW-tot2}
\end{center}
\end{figure} 

Finally, when aggregating the SPs' profits and the consumer surplus into the social welfare quantity, Fig.~\ref{fig:SW-tot2} shows that the variation of the aggregated consumer surplus dominates.

\section{Conclusions}\label{sec:conclusions}

We conclude that URLLC support provides a competitive advantage to an SP against a competitor SP that supports its service offering on eMBB, as long as the URLLC QoS constraint is not very stringent and the network investment is not expensive. On the other hand, the users of the URLLC SP are always better than the users of the eMBB SP, except for very expensive deployment, while the best situation for the whole user base occurs when the URLLC QoS constraint is stringent enough for causing zero profit for the URLLC SP. 

\bibliographystyle{IEEEtran}
\bibliography{bib}


\end{document}